\documentclass[doublecol]{epl2}

\def\be{\begin{equation}}
\def\ee{\end{equation}}
\def\bfi{\begin{figure}}
\def\efi{\end{figure}}
\def\bea{\begin{eqnarray}}
\def\eea{\end{eqnarray}}

\title{The grand canonical catastrophe as an instance of condensation of fluctuations}
\shorttitle{The grand canonical catastrophe as an instance of condensation of fluctuations} 

\author{M. Zannetti\thanks{\email{mrc.zannetti@gmail.com}}}
\shortauthor{M. Zannetti}

\institute{
Dipartimento di Fisica {\it E.Caianiello} and CNISM, Unit\`a di Salerno,
Universit\`a di Salerno, via Giovanni Paolo II, 84084 Fisciano (SA), Italy 
}

\abstract{
The so-called grand canonical catastrophe of the density fluctuations in
the ideal Bose gas is shown to be a particular instance of the much more general
phenomenon of condensation of fluctuations, taking place in a large system, in or out
of equilibrium, when a single degree of freedom makes a macroscopic contribution to
the fluctuations of an extensive quantity. The pathological character of the
``catastrophe'' is demystified by emphasizing the connection between experimental
conditions and statistical ensembles, as demonstrated by the recent realization
of photon condensation under grand canonical conditions.}

\begin{document}

\maketitle

\section{Introduction}

The grand canonical behavior of the density fluctuations 
of an ideal Bose gas (IBG) is known 
to contradict the intuitive notion that fluctuations
ought to be suppressed by lowering the temperature. 
This is commonly referred to as the 
{\it grand canonical catastrophe}~\cite{Ziff,Fujiwara,Stringari},
alluding to a failure
of the grand canonical ensemble (GCE) for bosons in the condensed phase. 
Furthermore, the successful
realization of Bose-Einstein condensation (BEC) of ultracold atoms, which requires
the canonical or microcanonical framework, has contributed to consolidate the notion of naivety
and inadequacy of the GCE. Yet, the GCE results are admittedly exact.
Hence, the alleged flaws of the GCE do not arise from internal
inconsistencies, but from the supposedly unphysical nature of the conditions required
for the GCE to apply. 
However, with the recent experimental realization of BEC
in a gas of photons~\cite{Klaers},
complemented shortly after by the investigation of the condensate fluctuations~\cite{Schmitt},
the outlook has changed, because condensation in this case has been
achieved under grand canonical conditions. 
This is an important development with implications on the issue
of statistical ensembles, since the grand canonical catastrophe has been
proven to be a real physical effect.

In this paper we approach the fluctuation problem
in the framework of large deviation theory and we show
that the observation of the grand canonical catastrophe 
is of interest in a context much wider than the physics of bosons, as 
an instance of {\it condensation of fluctuations}~\cite{condensation,Jo,Gambassi}. 
Condensation of fluctuations is a phenomenon related to but distinct from
the usual condensation which appears after taking a thermodynamic {\it average}.
The latter, to be referred to as condensation on average, belongs to 
the realm of typical behavior, while the former normally is a rare event.
These different instances of condensation may appear jointly or disjointly.
Of particular interest is the case of non interacting systems in which
the distinction between the two becomes extreme and, therefore,
most clear since without interaction there cannot be condensation on
average and yet fluctuations may condense~\cite{Jo}.
In this respect, the IBG framework offers a rich and flexible scenario allowing to
cover both cases: of condensation of fluctuations with and without condensation
on average. This paper aims to show that when the two types of condensation occur simultaneously,
the grand canonical catastrophe, rather than being a pathology, 
is the observable manifestation of condensation of fluctuations.

In order to explain what condensation of fluctuations is about,
it is convenient to lay down the basic probabilistic
structure: Consider a generic
system with microscopic states $\omega$ 
in a macrostate described by the statistical ensemble $P(\omega|J,V)$, where $J$ stands
for a set of control parameters and $V$ for the size of the system. The
setting is very general: $\omega$ could be a single event in sample space as well as a
trajectory, in which case $V$ would involve the trajectory's time length. Here, since
we have in mind an equilibrium application, $V$ will be taken as the system's volume. 
Referring to this ensemble as the prior, let ${\cal M}(\omega)$
be an extensive observable which scales like $V$. Now, the probability that 
${\cal M}(\omega)$ takes the value $M$ is given by 
\be
P(M|J,V) =  \sum_{\omega} P(\omega|J,V) \delta({\cal M} - M). 
\label{II.1}
\ee
Though straightforward, this turns out to be a statement rich of consequences
since a relation is established between
the observation of a fluctuation in the prior and the imposition of a constraint on the 
same system. In fact, using terminology from statistical mechanics, 
$P(M|J,V)$ plays also the role of the partition function in the
constrained ensemble
\be
P_{\rm c}(\omega|M,J,V) = \frac{1}{P(M|J,V)} P(\omega|J,V) \delta({\cal M} - M),
\label{II.3}
\ee 
obtained by conditioning the prior to the event ${\cal M}(\omega)=M$.
If ${\cal M}(\omega)$ obeys a large deviation principle, the above
duality between fluctuations and constraints is reflected into the twofold role of
the rate function 
\be
I(m|J) = -\lim_{V \to \infty} \frac{1}{V} \ln P(M|J,V), \;\; m=M/V
\label{II.1ter}
\ee
as the entity controlling fluctuations in the prior and as
the free energy density in the constrained ensemble~\cite{Touchette}.

The formal framework is completed by the important observation that
the task of computing the rate function can be simplified resorting to a soft version of the constraint
(for details see~\cite{Touchette} or~\cite{Jo}), namely
by applying an exponential bias on the prior in place of the hard
$\delta$ function constraint. This yields the biased ensemble
\be
P_{\rm b}(\omega|s,J,V) = \frac{1}{K(s,J,V)} P(\omega|J,V) e^{s{\cal M}(\omega)}
\label{tilted.1}
\ee
where $K(s,J,V)= \sum_{\omega} P(\omega|J,V)e^{s{\cal M}(\omega)}$ is the 
biased partition function. Continuing with the language of statistical mechanics, the constrained
ensemble~(\ref{II.3}) describes the system enclosed by walls isolating with respect to ${\cal M}$,
while the biased ensemble describes the system in contact with an ${\cal M}$-reservoir.
Then, the biased free energy $Y(s,J)= \lim_{V \to \infty} \frac{1}{V}\ln K(s,J,V)$ and
$I(m|J)$ are related by the Legendre transformation
\be
I(m|J) = s^*m - Y(s^*,J)
\label{LF.1}
\ee
where $s^*$ is the biasing field such that
\be
\langle {\cal M} \rangle_{s^*} = m
\label{LF.2}
\ee
and where $\langle \cdot \rangle_s$ stands for the average in the biased ensemble.
Summarizing, through $I(m|J)$ a link is established between fluctuations in the
prior and typical behavior either in the constrained $P_{\rm c}(\omega|M,J,V)$ or in the biased ensemble
$P_{\rm b}(\omega|s^*,J,V)$, which are equivalent in the sense of ensemble theory.

The implication of this duality is that rare fluctuations, difficult
to observe, can be made typical by the implementation of constraints~\cite{duality}.  
Conversely, if constraints are difficult or even impossible to realize in practice, in principle
constrained system could be studied through the fluctuations of 
unconstrained ones. The key point, for what follows, 
is that fluctuations in the unconstrained system
explore the phase diagram of the constrained one.

The probabilistic nature of the above structure makes it of wide applicability.
Among the many applications, as anticipated above, of particular interest 
is the apparently trivial case
of an uncorrelated prior. In fact, 
the imposition of a constraint
induces correlations which, in turn, may cause phase transitions to occur
in the constrained system. Perhaps,
the best known example of this type is the ferromagnetic transition due to the 
spherical constraint imposed on the Gaussian model~\cite{BK}. Then, as a consequence
of duality, the transition singularity appears also in the behavior of 
the prior's fluctuations while the prior's typical behavior, due to the absence of correlations, 
is smooth and featureless by construction.
In this case there will be condensation of fluctuations without condensation
on average. This phenomenon
has been studied in different contexts such as information theory~\cite{Merhav},
finance~\cite{Marsili} and statistical mechanics~\cite{condensation,Jo,Gambassi},
encompassing both equilibrium and out of equilibrium behavior.

\section{Grand canonical vs mean canonical ensemble}

The rest of the paper is devoted to the study of the 
implications of the general structure expounded above in the context of the IBG in equilibrium.
Focus will be on the boson number fluctuations. 
In order to make the presentation as simple as possible,
a uniform system in a $d$-dimensional box of volume $V$
with periodic boundary conditions will be considered.
The microstates $\omega$ are the sets of occupation numbers $\omega = \{ n_{\vec p} \}$ of the 
single particle momentum eigenstates.
The energy function is separable
${\cal H}(\omega) = \sum_{\vec p} n_{\vec p}\epsilon_p$ and
the single particle dispersion relation is of the form $\epsilon_p = ap^{\alpha}$,
where $a$ is a proportionality constant. For instance, for photons $a=c$ velocity of light and 
$\alpha = 1$, while
for particles with mass $\mathsf{m}$, $a=1/(2 \mathsf{m})$ and $\alpha = 2$.

The statistical ensemble is determined by the system's preparation protocol. Thus, in the GCE
which applies when the system is put in contact with a thermal and a particle reservoir,
the probability of a microstate is given by
\be
P_{\rm gc}(\omega|\beta,\mu,V) = \frac{1}{Z_{\rm gc}(\beta,\mu,V)} 
e^{-\beta [ {\cal H}(\omega)-\mu{\cal N}(\omega)]},
\label{IBG.50}
\ee
where ${\cal N}(\omega) = \sum_{\vec p} n_{\vec p}$ is the number function.
For large enough $V$, the equation of state takes the form~\cite{Huang}  
\be
\rho  =  \frac{1}{V} \frac{z}{1-z} + \lambda^{-d} g_{\nu}(z),
\label{pr.3}
\ee
where $\rho$ is the average density, $z=e^{\beta \mu}$
is the fugacity, $\lambda$
is the thermal wavelength~\cite{Yan}, 
\be
g_{\nu}(z) = \frac{1}{\Gamma(\nu)} \int_0^{\infty} dx \, \frac{x^{\nu-1}z}{e^{x} -z}
\label{pr.5}
\ee
is the Bose function with $\nu=d/\alpha$ and $\Gamma$ is the Euler gamma function.
In the above formula the Bose function
collects the contributions to the density from the excited states, 
while the $V$-dependent term is due to the average occupation of the ground state.
In the $\nu >1$ case, to which we will restrict from now
on, $g_{\nu}(1)=\zeta(\nu)$, where $\zeta$ is the Riemann zeta function.
Hence, in the $V \rightarrow \infty$ limit the isotherms become superiorly bounded by 
the critical value
\be
\rho_C(\beta) = \lambda^{-d} g_{\nu}(1).
\label{pr.7}
\ee
This means that, by taking the thermodynamic limit {\it and} adopting the GCE protocol, it
is not possible to prepare the system with an average density above 
$\rho_C$.  In other words, BEC does not take place in the GCE, as it is
well known from the example of black body radiation, corresponding to the GCE with $z=1$.
Occurrence of BEC, as condensation on average, requires number conservation which can
be implemented either rigidly or softly. In the first case, the number of particles
is fixed and the ensemble statistics is canonical
\be
P_{\rm c}(\omega|\beta,N,V) = \frac{1}{Z_{\rm c}(\beta,N,V)} e^{-\beta{\cal H}(\omega)} \delta_{{\cal N},N}.
\label{can.1}
\ee
In the second case the number of particles is fixed
on average~\cite{Huang} by inverting the density-fugacity relation. This amounts to introduce
the new ensemble defined by
\be
P_{\rm mc}(\omega|\beta,\rho,V) = P_{\rm gc}(\omega|\beta,z^*,V), 
\label{new.1}
\ee
where the function $z^*(\beta,\rho,V)$ is the unique root of Eq.~(\ref{pr.3}) given by (see Appendix 1)
\be
\ln z^*(\beta,\rho,V)  = \left \{ \begin{array}{ll}
         \ln g_{\nu}^{-1}(\lambda^d \rho),\;\; $for$ \;\; 
\rho < \rho_C   ,\\
         -AV^{-1/\nu},\;\; $for$ \;\; \rho = \rho_C  ,\\
         -1/[V(\rho-\rho_C)]  ,\;\; $for$ \;\; \rho > \rho_C,  
        \end{array}
        \right .
        \label{bec.10}
        \ee
with $A=[-\lambda^d \Gamma(\nu) \sin(\pi \nu)/\pi]^{1/\nu}$. 
In the following, GCE will denote only
the ensemble controlled by $(\beta,z,V)$, while the ensemble controlled by
$(\beta,\rho,V)$ will be referred to as the {\it mean canonical} ensemble (MCE), adopting
terminology from the spherical model literature~\cite{spherical}.
Though formally similar, these ensembles become non equivalent~\cite{equiv} in 
the thermodynamic limit: Dividing the density axis into the normal
phase below $\rho_C$ and the condensed phase above
$\rho_C$, the GCE and MCE overlap in the first one but not in the second,
which is accessible to the average density only in the MCE.

It should also be noted that
the origin of BEC in the MCE is in the 
mean-field treatment of the correlations generated by number conservation,
with $\mu^*=\ln z^*$
playing a role akin to that of the internal Weiss field in the mean-field theory
of ferromagnetism. Once this is understood, the distinction between MCE and GCE becomes
quite clear, since in the first ensemble the absence of interaction is  formal, while
the second one is genuinely non-interacting. This is the same distinction
existing between the mean spherical model and the Gaussian model of magnetic systems~\cite{BK,spherical}.
The similarities between the spherical model and the IBG have been pointed out by Kac and Thompson
in the paper of Ref.~\cite{spherical}. 
On the experimental  side, the control on the boson number is one of the key
elements in the realization of BEC in the laboratory. In experiments with
ultracold atoms condensation is achieved in (micro)canonical conditions
with control parameters $(\beta,N,V)$ by enclosing a definite number of atoms 
in optical traps~\cite{Stringari}. Instead, in the 
work of Klaers et al.~\cite{Klaers},
photons condensation has been obtained operating with the MCE protocol, that is
by keeping fixed the average density $\rho$~\cite{Muller}.

Therefore, it is of interest to study the density fluctuations in the two ensembles, since on the
basis of the considerations made in the Introduction, in the GCE (or in the
MCE normal phase) condensation
of fluctuations without condensation on average is expected, while in the MCE condensed phase
the two types of condensation are expected to appear simultaneously.

\begin{figure*}
\centering
\includegraphics[width=0.7\textwidth]{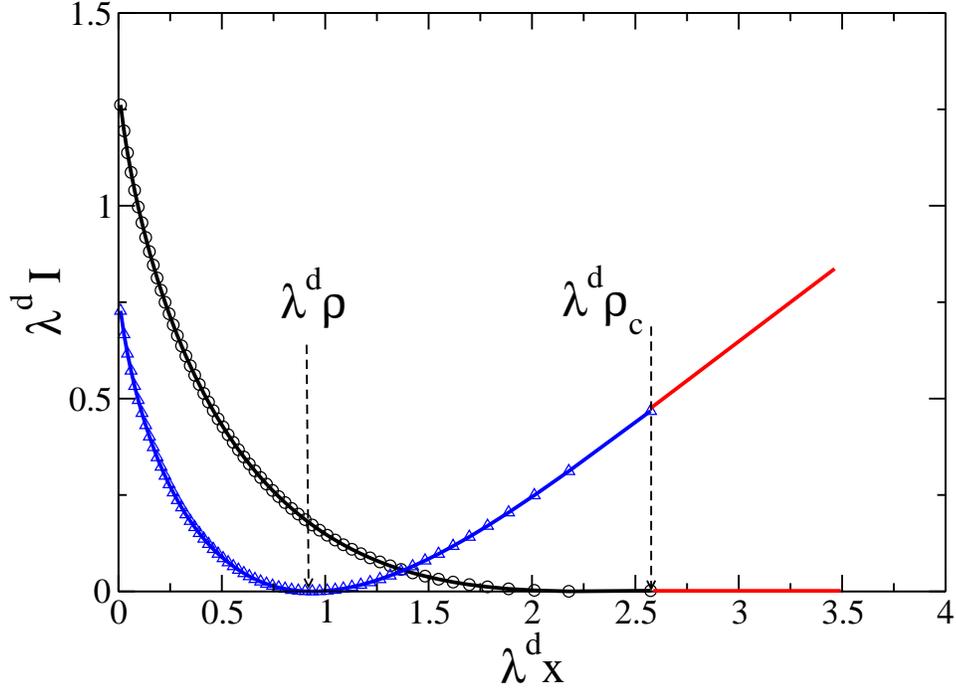}
\caption{Rescaled large deviation function vs rescaled fluctuating density
in the MCE: normal phase (blue triangles) with $-\ln z^* = 0.4$ and $\nu=3/2$,
condensed phase (black circles) with $\rho > \rho_C$ (Color on line).
It is a straightforward computation to show that the temperature dependence is rescaled away 
by the $\lambda^d$ factor.}
\label{fig.1}
\end{figure*}

\section{Density fluctuations}

As a consequence of the equivalence of the two ensembles in the normal phase, 
the study of fluctuations can be unified and carried out in the MCE framework for 
both of them.

Using Eq.~(\ref{new.1}),
the probability to find the value $N$ of ${\cal N}$ for a given $\rho$ reads
\be
P_{\rm mc}(N|\beta,\rho,V) =   \sum_{\omega} P_{\rm gc}(\omega|\beta,z^*,V)
\delta_{{\cal N},N}. 
\label{bec.4}
\ee
We are interested in checking whether ${\cal N}$ obeys a large deviation principle and,
if so, to find the rate function. This requires to extract the $V$-independent part
from ${\cal I}_{\rm mc}=-\frac{1}{V} \ln P_{\rm mc}$, evaluated for large $V$.
Carrying out the algebra expounded in Appendix 2 and denoting the fluctuating density
by $x=N/V$, for $x \leq \rho_C$ one obtains
\begin{eqnarray}
{\cal I}_{\rm mc}(x|\beta,\rho,V) & = & x \ln \frac{z^*(x)}{z^*(\rho)}  + \frac{1}{V} \ln z^*(x)   
\nonumber \\
& + & \frac{1}{V} \ln \frac{Z_{\rm gc}(\beta,z^*(\rho),V)}{Z_{\rm gc}(\beta,z^*(x),V)},
\label{calI.1}
\end{eqnarray}
with the function $z^*$ defined by Eq.~(\ref{bec.10}).
Instead, for $x > \rho_C$ one finds the linear behavior
\be
{\cal I}_{\rm mc}(x|\beta,\rho,V) = (\rho_C-x)\ln z^*(\rho) + {\cal I}_{\rm mc}(\rho_C|\beta,\rho,V).
\label{calI.2}
\ee
Here, two comments are in order. The first one is about duality, which appears in a simplified 
form in Eq.~(\ref{calI.1})
since ${\cal I}_{\rm mc}$ is related to the partition function in the {\it same} ensemble 
$Z_{\rm mc}(\beta,x,V)= Z_{\rm gc}(\beta,z^*(x),V)$. The reason is that by taking the MCE as prior
the bias
needed to render typical the fluctuation $x$ amounts just to shift the fugacity from
$z^*(\rho)$ to $z^*(x)$, without changing the form of the ensemble. 
Hence, with respect to number fluctuations,
the MCE is {\it self-dual}.
The second one is that the linear branch of Eq.~(\ref{calI.2})
corresponds to condensation of fluctuations.
In fact, exponentiating and comparing with the probability of the ground state occupation number
\be
P_{0}(n_0|\beta,z,V) = (1-z)e^{n_0\ln z},
\label{nzero.1}
\ee
there follows that for $x > \rho_C$ and up to normalization one can write
\be
P_{\rm mc}(N|\beta,\rho,V) = P_{0}(N-N_C|\beta,z^*(\rho),V)P_{\rm mc}(N_C|\beta,\rho,V),
\label{bec.43}
\ee
which shows
that macroscopic fluctuations above threshold can occur {\it only}
through the zero momentum occupation number, while the contribution
of the excited states is locked to $N_C=V\rho_C$.

The next step is to analyze separately the above result when
$\rho$ is in the normal or in the condensed phase. We shall see that
the $V \to \infty$ limit yields different behaviors in the two cases.

\subsection{Normal phase}

As follows from Eq.~(\ref{bec.10}), in the normal phase $z^*(\rho)<1$.  Furthermore, $z^*(\rho)$
is independent of $V$. Thus, letting $V \to \infty$,
there exists the rate function 
\begin{eqnarray}
& & I_{\rm mc}(x|\beta,\rho) =  \lim_{V \to \infty}{ \cal I}_{\rm mc}(\rho|\beta,\rho,V)  \nonumber \\ 
& = & \left \{ \begin{array}{ll}
x \ln \frac{z^*(x)}{z^*(\rho)} + \beta[f_{\rm mc}(\beta,x) - f_{\rm mc}(\beta,\rho)],
x \leq \rho_C \\
(\rho_C - x) \ln z^*(\rho) + I_{\rm mc}(\rho_C|\beta,z^*(\rho)),
\;\; x > \rho_C, 
        \end{array}
        \right .
        \label{K.6bis}
        \end{eqnarray}
where the MCE free energy density is given by~\cite{Huang}
$f_{\rm mc}(\beta,\rho) = -\beta^{-1}\lambda^{-d}g_{\nu+1}(z^*(\rho))$. As the plot 
displayed in Fig.~\ref{fig.1} (curve with
triangles) shows, this is a convex function with an isolated minimum 
at the typical value $\rho$,
where it vanishes, which is strictly convex for $x \leq \rho_C$ and
linear for $x > \rho_C$.

Hence, in the normal phase of the MCE (or
in the GCE) when $x$ exceeds the critical threshold
fluctuations do condense in absence of condensation on average. 
Though interesting, this fluctuation
phenomenon is doomed to remain an hardly observable event, since it is
exponentially suppressed for large $V$. In fact, due to the existence of the isolated minimum at $\rho$
\be
\lim_{V \to \infty} P_{\rm mc}(x|\beta,\rho,V) = \delta(x -\rho).
\label{ratef.2}
\ee
As we shall see, this situation is drastically changed in the condensed phase 
where condensation of fluctuations coexists with condensation on average.

%
%

\subsection{Condensed phase}

When $\rho$ is in the condensed phase, from Eq.~(\ref{bec.10}) follows that $z^*(\rho)$ becomes
$V$ dependent with $\lim_{V \to \infty}z^*(\rho)=1$. Hence, from Eqs.~(\ref{calI.1}) and~~(\ref{calI.2})
follows
\begin{eqnarray}
& & I_{\rm mc}(x|\beta,\rho)  = \nonumber \\
& & \left \{ \begin{array}{ll}
x \ln z^*(x) + \beta[f_{\rm mc}(\beta,x) - f_{\rm mc}(\beta,\rho_C)], x \leq \rho_C   ,\\
        0, \; x > \rho_C
        \end{array}
        \right .
        \label{mce.16}
        \end{eqnarray}
which shows that the large deviation function vanishes identically for fluctuations above 
threshold~\cite{Gambassi} (see Fig.~\ref{fig.1}).
This means that when both $\rho$ and $x$ are in the condensed phase, 
fluctuations behave sub-exponentially with respect to
$V$ and that in order to get the probability of fluctuations $V$ dependent terms must
be retained. Putting together the first line of Eq.~(\ref{mce.16}) and lowest order
terms from Eq.~(\ref{calI.2}) one has
\be
{\cal I}_{\rm mc}(x|\beta,\rho,V) = \left \{ \begin{array}{ll}
I_{\rm mc}(x|\beta,\rho), x \leq \rho_C \\
\frac{1}{V}\left [ \left (\frac{x-\rho_C}{\rho-\rho_C} \right) + \ln(V(\rho-\rho_C)) \right],
x > \rho_C.
\end{array}
\right.
\label{calI.1bis}
\ee  
Hence, due to the $1/V$ factor in the second line of the above equation, 
the probability of a fluctuation for $x > \rho_C$ becomes $V$ independent and,
in the $V \to \infty$ limit, instead of the $\delta$ function of Eq.~(\ref{ratef.2})
now one obtains the distribution
\be
P_{\rm mc}(x|\beta,\rho) 
= \left \{ \begin{array}{ll}
0   ,\;\; 
$for$ \;\; x \leq \rho_C, \\
\frac{\exp\{-\frac{x - \rho_C}{\rho - \rho_C}\}}
{(\rho - \rho_C)}, \;\; 
$for$ \;\; x > \rho_C.
\end{array}
        \right .
        \label{mce.17}
        \ee
which can be recognized as the Kac function of Ref.~\cite{Ziff} and which is 
also the statement of condensation of fluctuations,
as it can be checked by comparison with
Eq.~(\ref{nzero.1}). This is the main result in the paper.

With such a broad
distribution the distinction between typical and rare events gets blurred and the
width of fluctuations becomes macroscopic, since it goes like the density of the condensate
\be
\sqrt{\langle x^2 \rangle -\langle x \rangle^2} =  \rho - \rho_C.
\label{mce.18}
        \ee
Such a regime of strong fluctuations persists and is even enhanced by lowering
the temperature, since $\rho_C$ vanishes as $\beta \rightarrow \infty$. 
Therefore,  these features, which constitute the grand canonical catastrophe, are nothing but
the manifestation of condensation of fluctuations which, in the condensed phase, is as
typical as condensation on average.

\section{Conclusions}

The possibility to observe such a strong fluctuation regime depends only
on the realization of the physical conditions specific of the MCE. 
Even though the experiment by Klaers et al.~\cite{Klaers} is modeled
by bosons trapped in an harmonic potential, yet the conceptual structure expounded above still holds
suggesting that the large fluctuations of the condensate reported
in Ref.~\cite{Schmitt} are due to condensation of fluctuations. 

Summarizing, density fluctuations of the ideal Bose gas in thermal
equilibrium have been analyzed in the normal phase (which is equivalent to GCE) and in the
condensed phase of the MCE. In the first case  
condensation of fluctuations takes place as a rare event and in absence of condensation on
average. In the second case the two types of condensation occur simultaneously, both
as typical events and producing the phenomenology of the grand canonical catastrophe, which,
therefore, should not be regarded as a pathology, but just as an observable feature of the MCE.
The natural question is whether the conditions responsible of these macroscopic fluctuations could
be realized in atomic systems. Typically, atoms confined in optical traps are in microcanonical
conditions and, therefore, the overall density does not fluctuate. 
However, one could think to have such a large system that a subsystem could be regarded
in a GCE with energy and particle reservoirs due to the rest of the system. Then, if the chemical
potential of the subsystem is controlled through the density $\rho_{\rm ext}$ of the outer
system, it was shown by Ziff et al.~\cite{Ziff} that  
$\lim_{V \to \infty} P(x|\beta,\rho_{\rm ext},V)=\delta (x-\rho_{\rm ext})$ for $\rho_{\rm ext}$ in the 
normal as well
as in the condensed phase. Hence, there would be no grand canonical catastrophe. However, it
has been pointed out by Schmitt et al.~\cite{Schmitt} that this prediction is not in conflict
with their findings, because it pertains to experimental conditions
in which the subsystem is in diffusive contact with the environment, which is not the case
for the photon experiment. The point is that the experiment of Schmitt et al. takes
place in genuine MCE conditions, meaning by this that the control parameter is the average density
$\rho$ of the system {\it itself}, as opposed to $\rho_{\rm ext}$ in the conditions envisaged by Ziff et al. 
Though the distinction may seem rather subtle, yet it is enough to produce different 
statistical ensembles
as demonstrated by the drastically different fluctuation regimes in the condensed phase.

\section{Appendix 1}

Derivation of Eq.~(\ref{bec.10}):
If $\rho < \rho_C$ the first term in the right 
hand side of Eq.~(\ref{pr.3})
can be neglected, trivially yielding the $V$ independent solution in the first line of
Eq.~(\ref{bec.10}).
Instead, if $\rho \geq \rho_C$ one has $\epsilon = -\ln z \ll 1$. Therefore,
using the expansions $z \simeq 1-\epsilon$ and~\cite{macleod}
\be 
g_{\nu}(z)  = g_{\nu}(1) + \Gamma(1-\nu)\epsilon^{\nu-1}
\label{SMA.1}
\ee
Eq.~(\ref{pr.3}) can be rewritten as
\be
\rho - \rho_C  = \frac{1}{V\epsilon}  + \frac{\Gamma(1-\nu)}{\lambda^d}\epsilon^{\nu-1},
\label{SMA.2}
\ee
from which, it is straightforward to obtain the second and third line of Eq.~(\ref{bec.10}), under the
assumption $\nu < 2$, i.e. $d < 2\alpha$, with 
$\Gamma(1-\nu) = \pi/[\Gamma(\nu)\sin(\pi \nu)]$.
Notice that the condition on $\nu$ is fulfilled for particles with
mass and with $d < 4$. In the experimental conditions of Refs.~\cite{Klaers,Schmitt} photons
do behave as particles with an effective positive mass.

\section{Appendix 2}

Using the integral representation of the Kronecker $\delta$
\be
\delta_{N, {\cal N}} = 
\oint_{\cal C} \frac{d z^{\prime}}{2\pi i} \, z^{\prime \, {\cal N}-N-1},
\label{Kro.1}
\ee
the probability~(\ref{bec.4}) takes the form
\be
P_{\rm mc}(N|\beta,\rho,V) = e^{-V\Phi(x|\beta,z^*(\rho),V)} \oint_{\cal C} \frac{d z^{\prime}}{2\pi i}  \,
\frac{1}{z^{\prime}} e^{V\Phi(x|\beta,z^{\prime},V)},
\label{pr.10}
\ee
where ${\cal C}$ is a complex contour enclosing the origin and 
\be
\Phi(x|\beta,z,V) = -x \ln z + \frac{1}{V} Z_{\rm gc}(\beta,z,V),
\label{pr.11}
\ee
with
\be
Z_{\rm gc}(\beta,z,V)= - \frac{1}{V} \ln (1-z) +  \lambda^{-d} g_{\nu+1}(z).
\label{pr.11bis}
\ee
Hence,
\begin{eqnarray}
{\cal I}_{\rm mc}(x|\beta,\rho,V) & = & -\frac{1}{V} \ln P_{\rm mc}(N|\beta,\rho,V) \nonumber \\
& = &  \Phi(x|\beta,z^*(\rho),V) + K(x|\beta,V),
\label{K.1}
\end{eqnarray}
where
\be
K(x|\beta,V) = -\frac{1}{V} \ln \oint_{\cal C} \frac{d z^{\prime}}{2\pi i}  \,
\frac{1}{z^{\prime}} e^{V\Phi(x|\beta,z^{\prime},V)}.
\label{K.2}
\ee
The saddle point of the integrand function is determined by 
$\frac{\partial }{\partial z^{\prime}}\Phi(x|\beta,z^{\prime},V) = 0$,
which yields the equation 
\be
x  = \frac{1}{V}\langle {\cal N} \rangle_{z^{\prime}},
\label{saddle.1}
\ee
whose solution $z^*(\beta,x,V)$, therefore, depends on $x$ according to Eq.~(\ref{bec.10}). 
Hence, $z^*(x)$ is independent of $V$ for
$x< \rho_C$ and $V$-dependent for $x \geq \rho_C$.
In the first case, from Eq.~(\ref{K.2}) follows straightforwardly
\be
K(x|\beta,V) = -\Phi(x|\beta,z^*(x),V) + \frac{1}{V} \ln z^*(x).
\label{K.3}
\ee
In the second one, since $z^*(x)= e^{-1/[V(x - \rho_C)]} \simeq 1$ 
with a very weak dependence on $x$, 
the integral is dominated by the region where the Bose function is well approximated by~\cite{macleod}
\be
g_{\nu + 1}(z^{\prime}) \simeq g_{\nu + 1}(1) + g_{\nu}(1)\ln z^{\prime}.
\label{rfunc.2}
\ee
As a check, one can verify that the saddle point equation obtained with the above substitution coincides,
in the region of interest, with Eq.~(\ref{saddle.1}). Then, using the above form of $g_{\nu + 1}(z^{\prime})$
the integral can be carried out exactly, yielding
\be
K(x|\beta,V) = -\lambda^{-d} g_{\nu + 1}(1),
\label{K.4}
\ee
which is independent of $x$. Inserting these results into Eq.~(\ref{K.1}), one finds
\begin{eqnarray}
& & {\cal I}_{\rm mc}(x|\beta,\rho,V)  = \nonumber \\
& & \Phi(x|\beta,z^*(\rho),V) - \Phi(x|\beta,z^*(x),V) + \frac{1}{V} \ln z^*(x), \nonumber \\
\label{K.5}
\end{eqnarray}
for $x < \rho_C$ and
\be
{\cal I}_{\rm mc}(x|\beta,\rho,V)  =   \Phi(x|\beta,z^*(\rho),V) -\lambda^{-d} g_{\nu + 1}(1),
\label{K.6}
\ee
for $x \geq \rho_C$, which coincide with Eqs.~(\ref{calI.1}) and~(\ref{calI.2}).


\begin{thebibliography}{0}








\bibitem{Ziff}
 \Name{Ziff R. M., Uhlenbeck G. E. \and Kac M.}
 \REVIEW{Phys. Rep.}{32}{1977}{169}.

\bibitem{Fujiwara}
 \Name{Fujiwara I., ter Haar D. \and Wergeland H.}
 \REVIEW{J. Stat. Phys.}{2}{1970}{329};
 \Name{Holthaus  M. \and Kirsten K.}
 \REVIEW{Ann. of Phys.}{270}{1998}{198};
 \Name{Kocharovsky Vitaly V. V., Kocharovsky Vladimir V., Holthaus M., Raymond Ooi C. H., 
Svidzinsky A. A., Ketterle W. \and Scully M. O.}
 \REVIEW{Adv. in Atomic, Molecular and Opt. Phys.}{53}{2006}{291}.


\bibitem{Stringari}
\Name{Pitaevskii L. \and Stringari S.}
  \Book{Bose-Einstein Condensation}
  \Publ{Oxford University Press, New York}
  \Year{2003}.



\bibitem{Klaers}
 \Name{Klaers J., Schmitt J., Vewinger F. \and Weitz M.}
 \REVIEW{Nature}{468}{2010}{545}.

\bibitem{Schmitt}
 \Name{Schmitt J.,Damm T., Dung D., Vewinger F., Klaers J. \and Weitz M.}
 \REVIEW{Phys. Rev. Lett.}{112}{2014}{030401}.





\bibitem{condensation}
 \Name{Harris R.J., R\'akos A. \and Schuetz G.M.}
 \REVIEW{J. Stat. Mech.: Theory and Experiment}{P08003}{2005};
 \Name{Chleboun P. \and S. Grosskinsky S.}
 \REVIEW{J. Stat. Phys.}{140}{2010}{846};
 \Name{Corberi F. \and Cugliandolo L. F.} 
 \REVIEW{J. Stat. Mech.: Theory and Experiment}{P11019}{2012};
 \Name{Szavits-Nossan J., Evans M. R. \and Majumdar S. N.}
 \REVIEW{Phys. Rev. Lett.}{112}{2014}{020602}
and \REVIEW{J. Phys. A: Math. Theor.}{47}{2014}{455004};
 \Name{Corberi F., Gonnella G., Piscitelli A. \and Zannetti M.}
 \REVIEW{J. Phys. A: Math. Theor.}{46}{2013}{042001};
 \Name{Ferretti L., Mamino M. \and Bianconi G.}
 \REVIEW{Phys. Rev. E}{89}{2014}{042810}.

\bibitem{Jo}
 \Name{Zannetti M., Corberi F. \and Gonnella G.}
 \REVIEW{Phys. Rev. E}{90}{2014}{012143}; 
 \Name{Zannetti M., Corberi F., Gonnella G. \and Piscitelli A.}
 \REVIEW{Commun. Theor. Phys.}{62}{2014}{555}, {arXiv:1409.2798v1}.


\bibitem{Gambassi}
 \Name{Gambassi A. and Silva A.} 
 \REVIEW{Phys. Rev. Lett.}{109}{2012}{250602}.


\bibitem{Touchette}
 \Name{Touchette H.} 
 \REVIEW{Phys. Rep.}{478}{2009}{1}.

\bibitem{duality}
 \Name{Ritort F.}
 \REVIEW{J. Stat. Mech.: Theory and Experiment}{P10016}{2004};
 \Name{Derrida B.} 
 \REVIEW{J. Stat. Mech.: Theory and Experiment}{P07023}{2007}; 
 \Name{Giardin\`a C., Kurchan J. \and Peliti L.} 
 \REVIEW{Phys. Rev. Lett.}{96}{2006}{120603}; 
 \Name{Giardin\`a C., Kurchan J., Lecomte V. \and Tailleur J.} 
 \REVIEW{J. Stat. Phys.}{145}{2011}{787};
 \Name{Nemoto T. \and Sasa S.} 
 \REVIEW{Phys. Rev. E}{84}{2011}{061113}{arXiv:1309.7200v2};
 \Name{Jack R. and Sollich P.}
 \REVIEW{Progr. Theor. Phys. Supp.}{184}{2010}{304}; 
 \Name{Chetrite R. and Touchette H.}
 \REVIEW{Phys. Rev. Lett.}{111}{2013}{120601}{ and arXiv:1405.5157}; 
\Name{Hurtado P. I. and Garrido P. L.} 
 \REVIEW{Phys. Rev. Lett.}{107}{2011}{180601};
 \Name{Espigares C. P., Garrido P. L. \and Hurtado P. I.}
 \REVIEW{Phys. Rev. E}{87}{2013}{032115}.

\bibitem{BK}
 \Name{Berlin T. H. \and Kac M.}
 \REVIEW{Phys. Rev.}{86}{1952}{821}.

\bibitem{Merhav}
 \Name{Merhav N. \and Kafri Y.}
 \REVIEW{J. Stat. Mech.: Theory and Experiment}{P02011}{2010}.

\bibitem{Marsili}
 \Name{Filiasi M., Livan G., Marsili M., Peressi M., Vesselli E. \and Zarinelli E.}
 arXiv:1201.2817v1; 
 \Name{Filiasi M., Zarinelli E., Vesselli E. \and Marsili M.}
 arXiv:1309.7795v1;
 










\bibitem{Huang}
\Name{K. Huang K.} 
\Book{Statistical Mechanics}
\Publ{John Wiley and Sons, New York}
\Year{1967};
\Name{Pathria R. K. \and Beale P. D.}, 
\Book{Statistical Mechanics 3d Edition} 
\Publ{Elsevier, Burlington}
\Year{2011} 

\bibitem{Yan}
 \Name{Yan Z.} 
 \REVIEW{Eur. J. Phys.}{21}{2000}{625}.



\bibitem{spherical}
In addition to Ref.~\cite{BK} see 
 \Name{Lewis H. W. \and Wannier G. H.}
 \REVIEW{Phys. Rev.}{88}{1952}{682};
 \Name{Kac M. and Thompson C. J.} 
 \REVIEW{J. Math. Phys.}{18}{1977}.



\bibitem{equiv}
 \Name{Touchette H.}, 
 \REVIEW{Europhys. Lett.}{96}{2011}{50010}.



\bibitem{Muller}
It was suggested by 
 \Name{M\"uller E. E.} 
 \REVIEW{Physica}{139A}{1986}{165} 
that BEC
of photons could be observable by resorting to a reflecting cavity, which
would change the GCE into a CE by controlling the photon number.


\bibitem{macleod}
 \Name{MacLeod A. J.} 
 \REVIEW{Computers in Physics}{11}{1997}{385}.




\end{thebibliography}
\end{document}